# Introduction to cyclotrons for medical applications

*A. Gerbershagen*
John Adams Institute, University of Oxford, Oxford, United Kingdom

**Abstract**
Cyclotrons provide compact, reliable beams for radionuclide production, particle therapy and accelerator research. Their apparent simplicity conceals a number of challenging design aspects: the RF system must remain synchronous with the circulating ions, the magnetic field must provide transverse stability while compensating the relativistic change of revolution frequency, and the injection and extraction systems must control losses at high beam power. This note develops the relevant relations with one consistent relativistic notation. It distinguishes the smooth-orbit formulae used for first estimates from the closed-orbit calculations required for an azimuthally varying field. Particular attention is given to RF acceleration, isochronous field shaping, alternating-gradient focusing, turn separation, and electrostatic or stripping extraction. The consequences for two principal medical applications are then identified: high-current beams at moderate energy for radionuclide production, and fixed-energy proton beams followed by downstream energy selection for radiation therapy.

## 1 Fundamental principle, development and medical roles

A cyclotron combines a static magnetic field with a radio-frequency (RF) electric field. The magnetic field bends the ion onto a sequence of approximately circular equilibrium orbits, whereas the RF field increases its energy whenever it crosses an accelerating gap. Lawrence proposed the cyclotron principle in 1930, following Widerøe’s earlier work on RF acceleration. Lawrence and Livingston operated the first 4.5-inch cyclotron in 1931 and reported the higher-energy 27-inch machine in 1932 [1, 2].

The first machines established the cyclotron as a practical research accelerator within only a few years. The 27-inch Berkeley cyclotron followed the early laboratory prototypes and reached energies of several megaelectronvolts (Fig. 1, left). Cyclotrons then became central instruments for nuclear-structure and reaction studies. From the 1980s onwards, compact machines for radionuclide production and particle therapy became an increasingly important part of the installed base. Superconducting magnets reduced the size of high-energy cyclotrons, while larger research facilities such as at GANIL, INFN Catania or PARTREC (Fig. 1, right) retained the flexibility to accelerate several ion species.

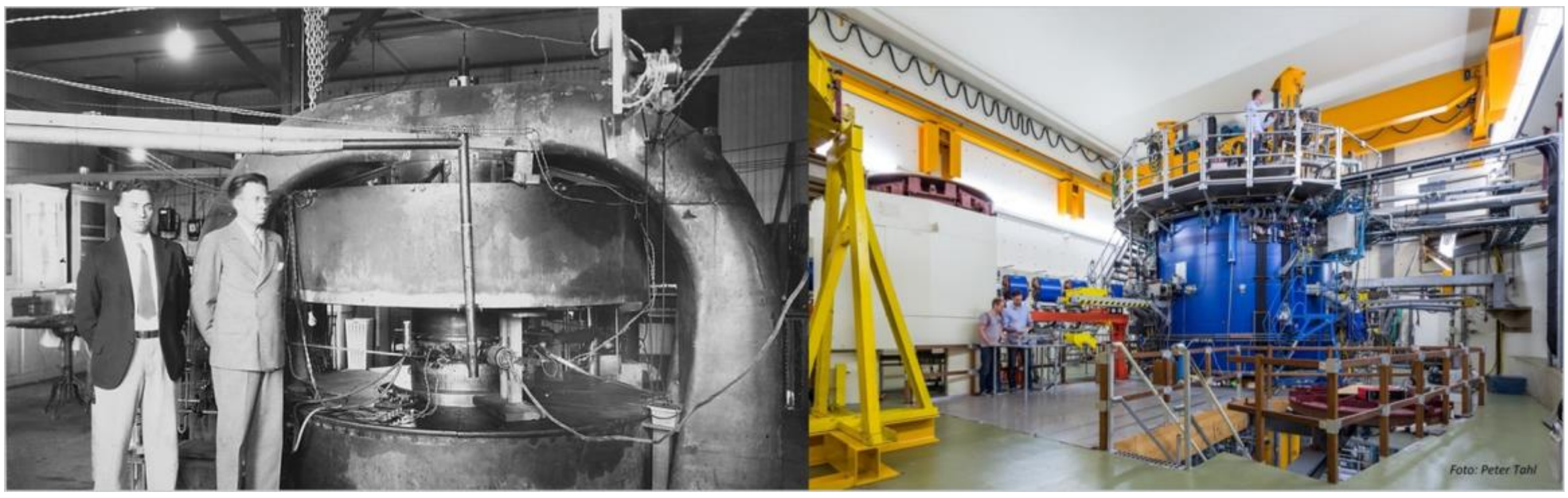

**Fig. 1:** Cyclotron development illustrated with photographs from lecture slides 6 and 7. Lawrence and Livingston stand in front of the 27-inch Berkeley cyclotron (left). The superconducting AGOR cyclotron is shown at Groningen (right).

Medical cyclotrons occupy two principal operating regimes [3, 4]. Compact positron-emission tomography cyclotrons commonly accelerate protons or $H^-$ ions to approximately 10–20 MeV. Larger radionuclide-production machines extend to several tens of megaelectronvolts and can reach milliampere-class current. The reaction $^{18}O(p,n)^{18}F$ is widely used to produce $^{18}F$ for positron-emission tomography [5]. In contrast, proton therapy requires roughly 70–250 MeV at the patient. A fixed-energy isochronous cyclotron commonly accelerates to 230–250 MeV. A variable degrader and an energy-selection system then set the treatment energy. The degrader introduces multiple scattering, energy straggling and nuclear losses. The downstream optics must therefore recover an acceptable emittance and momentum spread [6]. The IAEA and PTCOG maintain current facility directories for radionuclide-production and particle-therapy accelerators, respectively [10, 11].

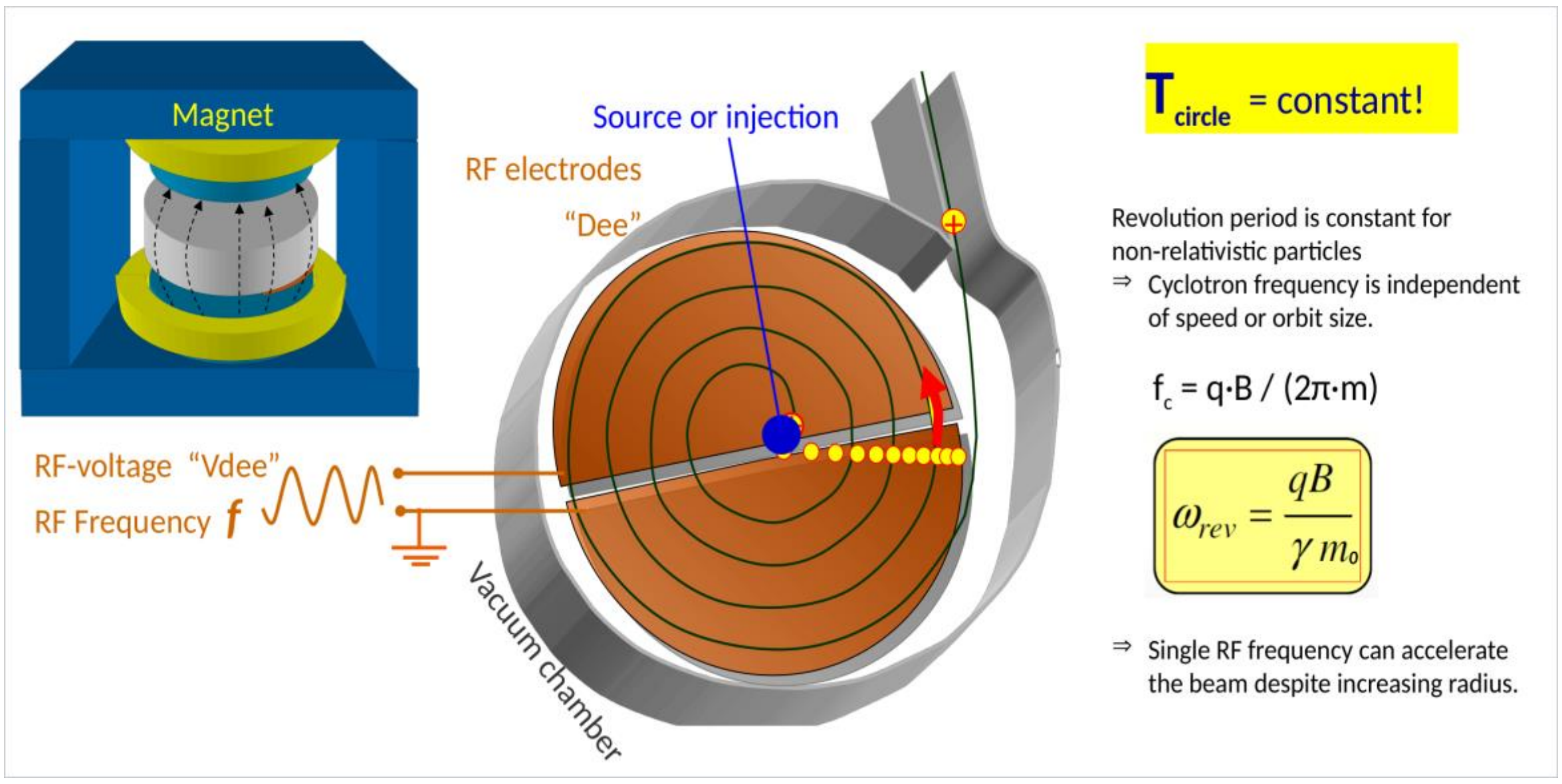

**Fig. 2:** Classical cyclotron principle. A static vertical magnetic field returns the ion to the accelerating gaps, while the orbit radius increases after each RF energy increment. The constant-period statement applies in the non-relativistic limit. The trajectory is schematic and does not represent the small turn separation of a real machine.

Figure 2 demonstrates the basic layout and the principle behind the cyclotron accelerator. Let $m_0$ denote the ion rest mass, $q$ its signed charge and $Q = |q|$ the charge magnitude. The kinetic energy $K$, momentum $p$, velocity $v = \beta c$ and Lorentz factor $\gamma$ are related by

$$K = (\gamma - 1)m_0c^2, \qquad p = \beta\gamma m_0 c, \qquad B\rho = \frac{p}{Q}. \tag{1}$$

Here $B\rho$ is the magnetic rigidity. In a uniform field, or in a smooth-orbit estimate using the mean field $\overline{B}$, the equilibrium condition $p = Q\overline{B}R$ gives

$$f_{\text{rev}} = \frac{Q\overline{B}}{2\pi\gamma m_0}. \tag{2}$$

The familiar non-relativistic result follows when $\gamma \simeq 1$. Equation (2) also identifies the central limitation of a uniform-field cyclotron. As the kinetic energy rises, $\gamma$ increases and the revolution frequency decreases. An isochronous cyclotron compensates this change through its radial field profile, whereas a synchrocyclotron varies the RF frequency during the acceleration cycle.

## 2 RF acceleration and longitudinal control

The beam gains energy only while crossing the gaps between conducting RF electrodes. The field inside a dee is small, so the particle drifts there at nearly constant energy. Modern dees may be sector-shaped or spiral rather than D-shaped. In a separated-sector cyclotron, dedicated cavities occupy the spaces between the sector magnets. Their larger volume permits a high accelerating voltage and a substantial energy gain per turn. The PSI High Intensity Proton Accelerator (HIPA) is the principal high-power example. Its ring cyclotron delivers protons at 590 MeV and up to 2.4 mA, corresponding to about 1.4 MW [9].

The RF frequency is selected as an integer harmonic $h$ of the revolution frequency,

$$f_{\text{RF}} = h f_{\text{rev}}. \tag{3}$$

The harmonic number must be consistent with the number and placement of accelerating gaps and with the orbit symmetry. If the ion crosses $N_g$ gaps during one turn, the energy increment is

$$\Delta K_{\text{turn}} = Q \sum_{j=1}^{N_g} V_{g,j}\, T_j \cos\phi_{s,j}. \tag{4}$$

In Eq. (4), $V_{g,j}$ is the peak voltage of gap $j$, $T_j$ is its transit-time factor and $\phi_{s,j}$ is the phase at which the bunch centre crosses the gap, with the crest defined by $\phi_s = 0$. A larger energy gain reduces the number of turns and normally increases the final-turn separation. It also reduces the time during which space charge and field errors act on the beam. These benefits must be balanced against cavity voltage, RF power, sparking margin and the radial space available for the resonators. Operation at harmonic $h$ can provide multiple equivalent RF phases around the machine. The number of phases that are populated, and the obtainable current, remain limited by injection, acceptance and space charge. A higher harmonic number does not by itself increase the current.

Residual revolution-frequency errors accumulate as RF phase error. For the $n$th turn, the exact phase increment relative to harmonic $h$ can be written as

$$\delta\phi_n = 2\pi \left[\frac{f_{\text{RF}}}{f_{\text{rev}}(K_n)} - h\right]. \tag{5}$$

If $f_{\text{rev}} = f_{\text{RF}}(1 + \epsilon_n)/h$ with $|\epsilon_n| \ll 1$, Eq. (5) reduces to $\delta\phi_n \simeq -2\pi h \epsilon_n$. Even a small systematic frequency error can therefore become significant after many turns. Field mapping, cavity-frequency control and beam-based phase measurements are used together to maintain the required tolerance.

The bunch occupies a finite RF phase interval, so the sinusoidal gap voltage produces a phase-dependent energy gain (see Fig. 3, left). A counter-phased third-harmonic cavity can cancel the quadratic curvature of the fundamental voltage at the crest (as illustrated on Fig. 3, right). The resulting waveform reduces the correlated energy spread of a bunch centred near the crest, but it does not correct injection-phase errors or turn-to-turn variations of the closed orbit. One convenient convention, for which $V''(0) = 0$ and the crest voltage is $8V_1/9$, is

$$V(\phi) = V_1 \left[\cos\phi - \frac{1}{9}\cos(3\phi)\right], \tag{6}$$

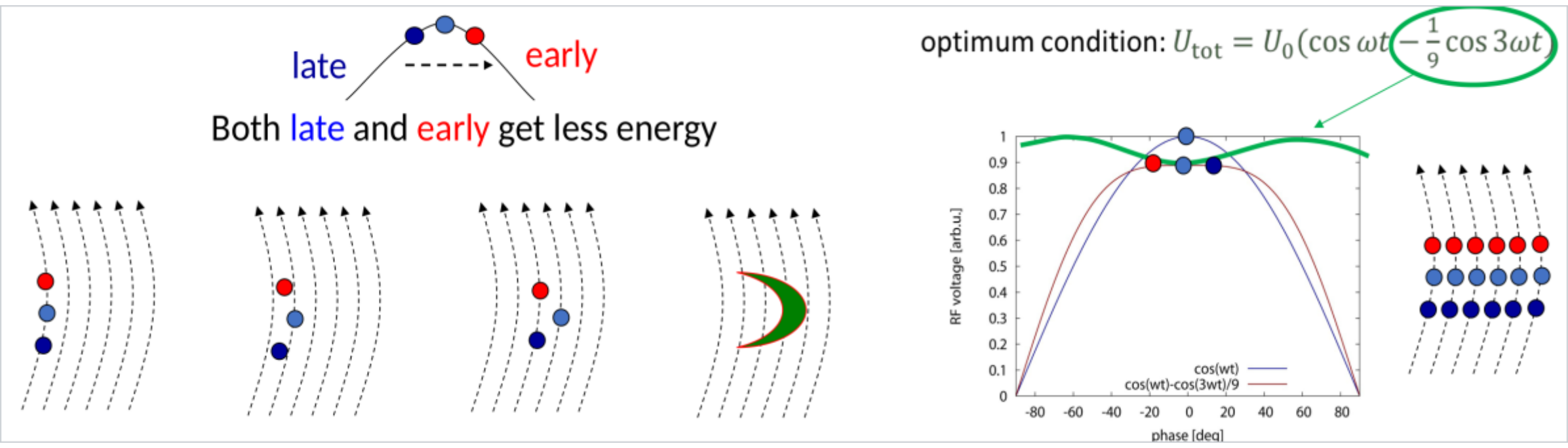

**Fig. 3:** Phase mixing and third-harmonic flattop correction. The panels are cropped and aligned from lecture slides 19 and 20. The left panel shows how phase-dependent energy gain produces radial broadening. The right panel shows the flatter voltage maximum obtained by adding a counter-phased third harmonic.

At high current the peak bunch current, rather than only the average current, determines the direct space-charge force. Longitudinal and transverse motion then become coupled. A stationary, approximately round bunch distribution, often described as vortex motion, can form in some compact high-intensity cyclotrons when space charge, external focusing and acceleration are appropriately balanced [3]. This behaviour is machine-dependent and operating-point-dependent. It must therefore be established with self-consistent simulations and measurements rather than assumed from a zero-current lattice.

Multipacting is a separate RF limitation. Electrons accelerated across an RF gap can release secondary electrons when they strike a metallic surface. An avalanche develops if the transit time is resonant with the RF field and the secondary-electron yield exceeds unity. The resulting current can load the cavity, increase heating and outgassing, and prevent the required voltage from being reached. Conditioning and surface treatment reduce the secondary yield. Geometry changes, bias fields and rapid passage through a multipacting barrier during RF ramp-up provide further mitigation. The appropriate method depends on the cavity and operating voltage.

## 3 Relativistic synchronism and field shaping

For fixed $f_{\mathrm{RF}}$ and $h$, Eq. (2) requires the mean field to rise in proportion to $\gamma$. In the smooth, axisymmetric approximation,

$$\overline{B}(R) = B_0\gamma(R), \qquad B_0 = \frac{2\pi m_0 f_{\mathrm{RF}}}{hQ}, \qquad R = R_\infty\beta, \quad R_\infty = \frac{m_0 c}{QB_0}. \tag{7}$$

Consequently, $\gamma = [1-(R/R_\infty)^2]^{-1/2}$. At 250 MeV, a proton has $\gamma = 1.266$, $p = 0.729$ GeV/$c$ and $B\rho = 2.43$ T m. The mean field of an idealised 250 MeV isochronous cyclotron must therefore rise by about 27% between the non-relativistic centre and extraction (see example in Fig. 4). Heavy ions place a stronger demand on magnetic rigidity: fully stripped carbon at 430 MeV/u has $B\rho \simeq 6.63$ T m. These values illustrate why compact therapy designs tend towards high magnetic field.

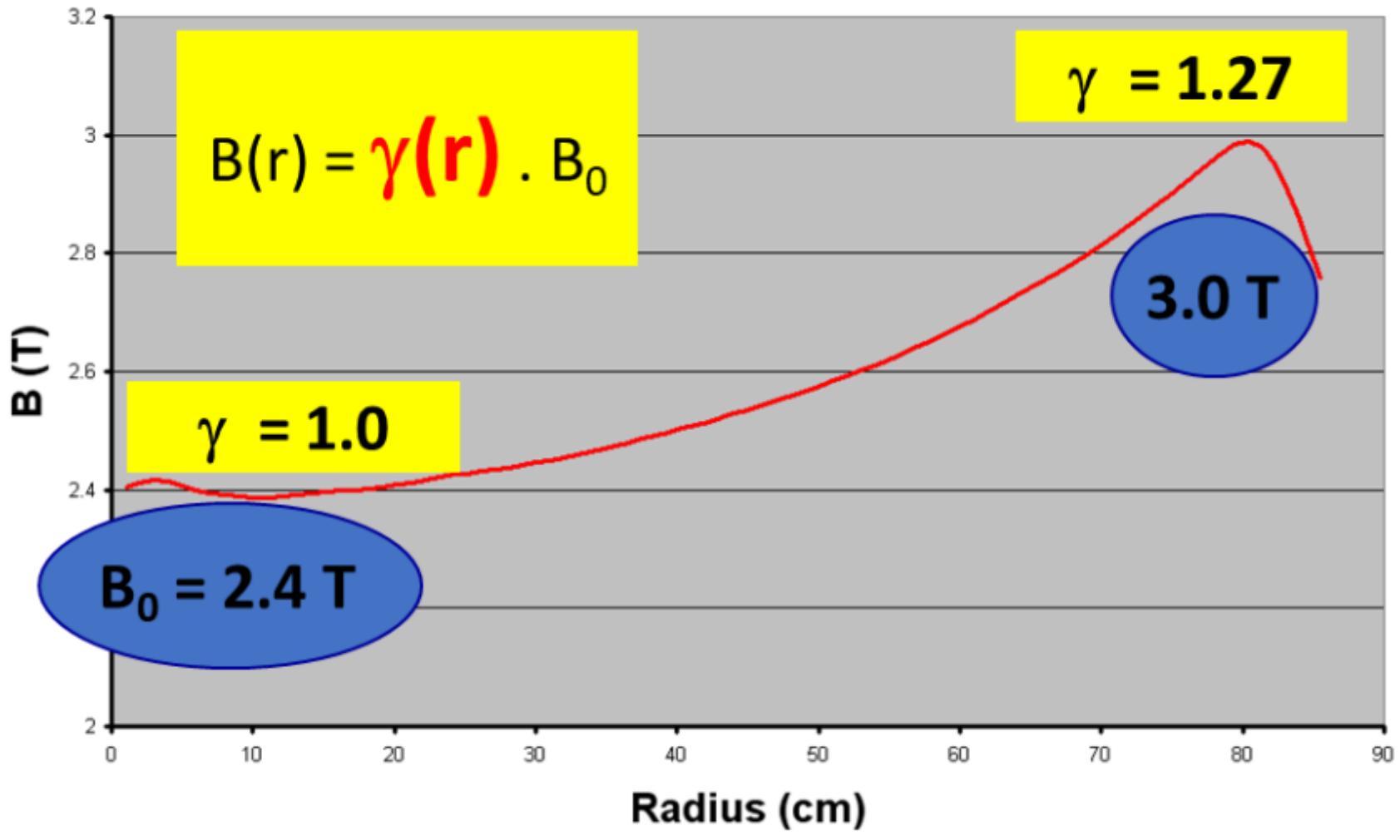

**Fig. 4:** Isochronous field shaping, demonstrated on the example of the orbit-averaged field of the COMET cyclotron, which rises with radius from approximately 2.4 T near the centre to about 3 T at extraction.

Equation (7) is a design relation, not an exact local-field prescription for an azimuthally varying field. Sector boundaries make the equilibrium orbit scallop, and the ion samples hills and valleys with unequal path length. The exact isochronism condition is therefore a constant closed-orbit time of flight. Three-dimensional magnetic-field calculations, numerical closed-orbit tracking and measured field maps are used to trim the pole profile and correction coils. The same calculation yields the orbit tunes and the locations of structural or imperfection resonances.

The required field increase can be produced in several ways. A decreasing pole gap raises the field towards the outer radius. Superconducting coils placed closer to the median plane can provide a similar radial dependence. Multiple main-coil systems and concentric trim coils offer greater flexibility when several ion species or extraction energies are required. The final profile is determined from the orbit-averaged field and time of flight, not from the local hill or valley field.

An isochronous cyclotron can instead accept repeated bunches and provide a quasi-continuous beam. This is advantageous for radionuclide production and high average power.

A synchrocyclotron makes the complementary choice of compensation for the increase in relativistic $\gamma$. Its magnetic field remains fixed while the RF frequency is swept downwards to follow the decreasing revolution frequency. One bunch, or a small number of bunches, is accelerated during each modulation cycle, so the extracted beam is pulsed.

## 4 Transverse focusing in an azimuthally varying field

Define the mean-field index

$$k = \frac{R}{\overline{B}}\frac{\mathrm{d}\overline{B}}{\mathrm{d}R}. \quad (8)$$

In a smooth weak-focusing model the radial and vertical tunes satisfy $\nu_r^2 \simeq 1 + k$ and $\nu_z^2 \simeq -k$. Isochronism requires $k > 0$ over most of the acceleration range. The corresponding smooth model is therefore vertically unstable. An azimuthally varying field (AVF) restores vertical focusing by alternating high-field hills and low-field valleys (see Fig. 5, left). Spiral sector boundaries strengthen this alternating-gradient action (see Fig. 5, right).

For a hard-edge sector model, define the field flutter and spiral angle by

$$F = \left\langle \left(\frac{B-\bar{B}}{\bar{B}}\right)^2 \right\rangle_\theta , \qquad \tan\xi = R\frac{\mathrm{d}\theta_{\mathrm{edge}}}{\mathrm{d}R}. \tag{9}$$

A useful leading-order estimate for $N > 2$ identical sectors is

$$\nu_z^2 \simeq -k + \frac{N^2}{N^2-1}F(1 + 2\tan^2\xi). \tag{10}$$

Equations (9) and (10) expose the design trade-off: larger flutter and spiral angle improve vertical focusing, whereas the required positive field index opposes it. The coefficients depend on the hard-edge and smooth-orbit approximations. Quantitative tune prediction for a real magnet must include the measured or computed three-dimensional field, finite sector edges and the scalloped closed orbit [3, 7].

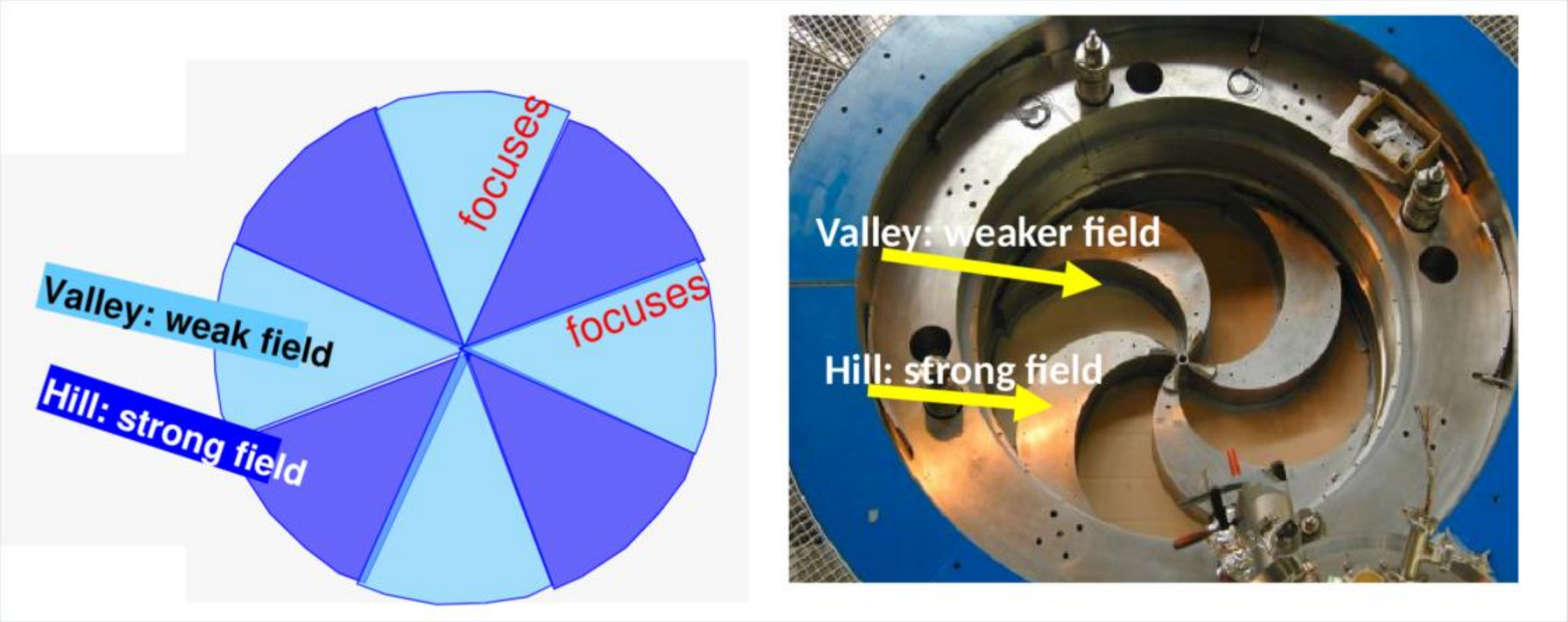

**Fig. 5:** Illustration of AVF and spiral-sector focusing. Alternating hill and valley fields provide focusing (left). The photograph shows the spiral pole geometry used to strengthen edge focusing at larger radius (right).

The working point must avoid resonances driven by the sector symmetry, field errors and extraction elements. Orbit centring is equally important. A coherent radial oscillation changes the apparent last-turn separation and can increase losses on an electrostatic septum. Central-region geometry, injection matching and trim coils are therefore adjusted using both orbit calculations and beam-based measurements.

## 5 Injection, turn separation and extraction

Compact isotope cyclotrons often place an ion source in the central region. In a Penning ion gauge source, a low-voltage arc discharge ionises the gas and the cyclotron field confines the electrons. The RF puller field extracts ions through the source aperture. Axial injection from an external source improves accessibility and uses a spiral inflector to bend the incoming beam into the median plane. Radial injection is mainly used in separated-sector cyclotrons, where a beam line can enter through a valley [8]. In every case, the first turns have small radius and tight geometric acceptance. The inflector voltage, RF phase, initial orbit centring and space-charge forces consequently determine the captured current.

Within the smooth-orbit approximation, differentiation of $p = Q\bar{B}R$ for neighbouring equilibrium orbits gives the first-order energy-driven radial step

$$\Delta R_{\mathrm{EO}} \simeq \frac{\Delta K_{\mathrm{turn}}}{m_0c^2}\frac{\gamma R}{(\gamma^2-1)(1+k)}. \tag{11}$$

In the ideal smooth isochronous case, $1 + k = \gamma^2$, so

$$\Delta R_{\rm iso} \simeq \frac{\Delta K_{\rm turn}}{m_0 c^2} \frac{R}{\gamma(\gamma^2-1)} \overset{K \ll m_0 c^2}{\rightarrow} \frac{R\, \Delta K_{\rm turn}}{2K}. \tag{12}$$

For illustration, $R = 0.8$ m, $K = 250$ MeV and $\Delta K_{\rm turn} = 0.5$ MeV give $\Delta R_{\rm iso} \simeq 0.56$ mm, which is smaller than the typical transverse beam size in the cyclotron, thus leading to the overlap of the beams that have made different number of turns in the machine. This is only the equilibrium-orbit contribution. The measured gap between the final beam turns also contains coherent radial motion, orbit distortion and the finite radial beam size. Numerical closed-orbit tracking is therefore required before a septum-loss estimate is credible. Increasing the final-turn voltage and arranging the extraction optics to exploit a controlled radial oscillation can enlarge the usable gap [8].

Positive-ion isochronous cyclotrons commonly use an electrostatic extraction system with a thin septum between the circulating and extracted turns (see Fig. 6, left and center) [8]. A deflector of transverse field $\mathcal{E}_d$ and effective length $L$ gives the approximate angular kick

$$\theta \simeq \frac{Q\mathcal{E}_d L}{pv} = \frac{\mathcal{E}_d L}{\mathcal{E}\rho}, \qquad \mathcal{E}\rho = \frac{pv}{Q}. \tag{13}$$

The electric rigidity $\mathcal{E}\rho$ rises with energy, so a high-energy deflector requires a long electrode, a high field, or both. For a 250 MeV proton, a field of 10 MV/m over 0.5 m gives a kick of about 11 mrad. Subsequent magnetic channels move the extracted orbit clear of the machine. The septum intercepts the halo and any overlap between adjacent turns (Fig. 6, right). Low emittance, large turn separation and accurate alignment are therefore essential for high-intensity beams.

An $H^-$ cyclotron can instead use stripping extraction. The foil removes the two electrons, so the proton retains nearly the same momentum but reverses its curvature in the same magnetic field. A movable foil can select the extraction radius and can provide more than one beam line. Stripping relaxes the requirement for a clean final-turn gap. Its limitations include foil scattering, incomplete interception, residual-gas stripping and the handling of stripped electrons. Foil heating and lifetime become important at high current. Synchrocyclotrons commonly use precessional or regenerative extraction. Precessional extraction uses a coherent first-harmonic radial oscillation as $\nu_r$ approaches unity. Regenerative extraction uses a second-harmonic gradient to enter the $2\nu_r = 2$ stopband and produces rapid amplitude growth.

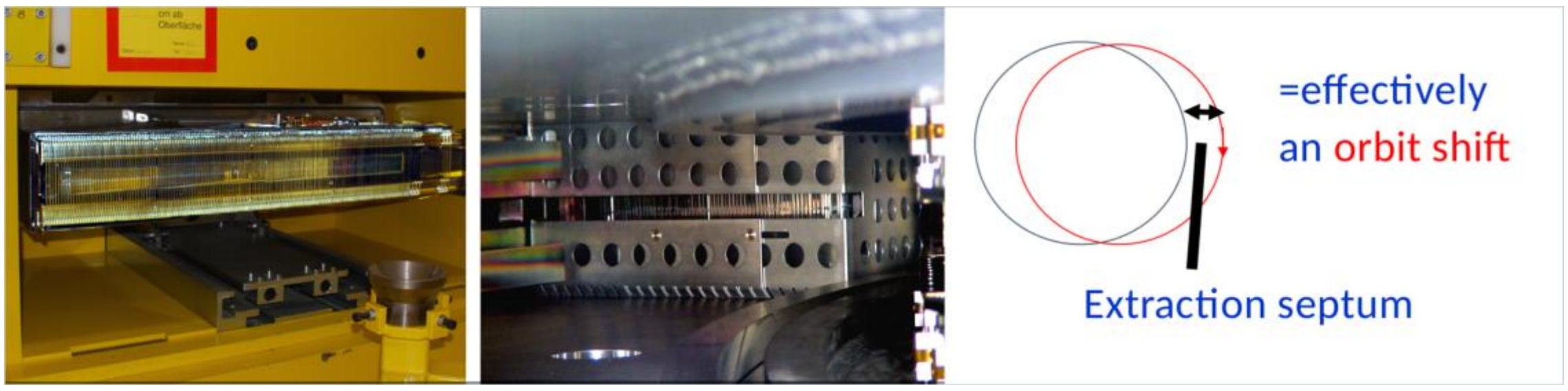

**Fig. 6:** The photographs show a high-voltage extraction channel and its narrow septum region (left and centre). A coherent radial oscillation can displace the final orbit outwards at the extraction septum (right).

Beam loss must be evaluated as power rather than only as fractional current. For a beam current $I$ defined as charge per unit time,

$$P_{\rm beam} = \frac{IK}{Q}, \qquad P_{\rm loss} = f_{\rm loss} P_{\rm beam}. \tag{14}$$

For protons, 1 mA at 1 MeV corresponds to 1 kW. The PSI High Intensity Proton Accelerator, for example, routinely operates at 590 MeV and up to 2.4 mA, corresponding to approximately 1.4 MW [9]. At this power a loss fraction of $10^{-4}$ still represents about 140 W. Collimation, shielding, component cooling and remote handling must therefore be treated as part of the accelerator design.

## 6 Conclusions

The cyclotron is a resonant accelerator that requires a coupling of RF synchronism, magnetic rigidity, transverse focusing and extraction geometry. Relativistic isochronism requires an increasing mean field. AVF sectors and spiral edges supply the vertical focusing lost in the corresponding smooth-field model. RF voltage controls both the acceleration time and the turn separation. Harmonic cavities can reduce correlated energy spread. At extraction, electrostatic systems demand well-separated, low-emittance turns.

These principles lead directly to the architecture of medical installations. Radionuclide production favours high average current at moderate energy and a robust target station. Proton therapy requires a substantially higher extraction energy and, for fixed-energy cyclotrons, losses in the degrader and energy-selection system dominate the downstream transmission and emittance. First-order formulae establish the design scales and expose the controlling parameters. Final performance, however, must be calculated with three-dimensional field models, self-consistent beam simulations, measured RF and magnetic fields, and beam-based commissioning data.